\global\def\draftcontrol{0}

   \def\versionno{ n=2star -- draft   }

\catcode`\@=11

\expandafter\ifx\csname draftcontrol\endcsname\relax\global\def\draftcontrol{0}
\fi

{\count255=\time\divide\count255 by 60
\xdef\hourmin{\number\count255}
\multiply\count255 by-60\advance\count255 by\time
\xdef\hourmin{\hourmin:\ifnum\count255<10 0\fi\the\count255}}
\def\draftdate{\number\month/\number\day/\number\year\ \ \ \hourmin }

\newcommand\makepapertitle{\par
  \begingroup
    \renewcommand\thefootnote{\@fnsymbol\c@footnote}%
    \def\@makefnmark{\rlap{\@textsuperscript{\normalfont\@thefnmark}}}%
    \long\def\@makefntext##1{\parindent 1em\noindent
            \hb@xt@1.8em{%
                \hss\@textsuperscript{\normalfont\@thefnmark}}##1}%
     \newpage
     \global\@topnum\z@   
     \@makepapertitle
     \thispagestyle{empty}\@thanks
  \endgroup
  \setcounter{footnote}{0}%
  \global\let\thanks\relax
  \global\let\makepapertitle\relax
  \global\let\@makepapertitle\relax
  \global\let\@thanks\@empty
  \global\let\@author\@empty
  \global\let\@date\@empty
  \global\let\@title\@empty
  \global\let\title\relax
  \global\let\author\relax
  \global\let\date\relax
  \global\let\and\relax
  \def\version{\let\version\@version\@gobble}
}
\def\@makepapertitle{%
  \newpage
   \ifnum\draftcontrol=1 {}
   \version\versionno
   \vskip 3em%
   \else
   \hfill\hbox to 3cm {\parbox{4cm}{\@pubnum}\hss}%
   \vskip 3em%
   \fi
   \begin{center}%
   \let \footnote \thanks
     {\LARGE {\@title}}%
     \vskip 1.5em%
     {\normalsize
       \lineskip .5em%
       \begin{tabular}[t]{c}%
         \@author
       \end{tabular}\par}%
     \vskip 1.5em%
     {\@bstract}%
     \end{center}%
     \vskip 1.5em
     \@date%
   \par
}

\gdef\@pubnum{}
\def\pubnum#1{%
  \gdef\@pubnum{#1}}

\gdef\@bstract{}
\def\Abstract#1{%
  \gdef\@bstract{%
   \parbox{\textwidth-0pc}{%
   \centerline{\bf Abstract}\penalty1000%
\kern.2cm%
\noindent
\renewcommand\baselinestretch{1.0}%
{#1}}}
}

\def\ps@paper{\let\@mkboth\@gobbletwo%
     \ifnum\draftcontrol=1
	\def\@oddfoot{\hbox to \textwidth{\tiny \versionno \hfil\tiny\draftdate}%
	\hskip -\textwidth \hbox to \textwidth{\hfil\rm\thepage\hfil}}%
     \else\def\@oddfoot{\hbox to \textwidth{\hfil\rm\thepage\hfil}}
     \fi
     \let\@evenfoot\@oddfoot
}

\def\body{\clearpage
          \pagestyle{paper}
	}

\def\@version#1{\ifnum\draftcontrol=1
\typeout{}\typeout{#1}\typeout{}
\vskip3mm\centerline{\hbox{\fbox{\normalsize{\tt DRAFT -- #1 -- }
                   {\draftdate}}}}\vskip3mm
\fi}
\let\version\@version
\long\def\eqlabel#1{\ifnum\draftcontrol=1
                    \tag@false  
                    \tag*{(\theequation) \hbox to -0.2cm{\hspace{0cm}\small{#1}\hss}}
                    \refstepcounter{equation}
                    \edef\@currentlabel{\theequation}
                    \ltx@label{#1}          
                    \else
                    \label{#1}
                    \fi
                    }
\let\st@bibitem\@bibitem
\let\st@lbibitem\@lbibitem
\ifnum\draftcontrol=1
  \def\@bibitem#1{%
    \st@bibitem{#1}\a@@label{#1}\ignorespaces}
  \def\@lbibitem[#1]#2{%
    \st@lbibitem[#1]{#2}\a@@label{#2}\ignorespaces}
  \def\a@@label#1{%
    \gdef\a@lab{\smash{\normalfont\small#1}}
    \ifvmode
      \if@inlabel
        \global\setbox\@labels\hbox{%
          \llap{\a@lab\let\a@lab\relax
                \kern\@totalleftmargin\kern\marginparsep}%
          \box\@labels}%
      \fi
    \fi}
\fi

\documentclass[12pt,letterpaper]{article}

\usepackage{amsmath,amssymb,array,calc,rotating,epsfig,psfrag}
\usepackage[nosort]{cite}

\ifnum\draftcontrol=1
\fi

\renewcommand\baselinestretch{1.25}
\renewcommand\section{\@startsection {section}{1}{\z@}%
                                   {-3.5ex \@plus -1ex \@minus -.2ex}%
                                   {2.3ex \@plus.2ex}%
                                   {\normalfont\large\bfseries}}
\renewcommand\subsection{\@startsection{subsection}{2}{\z@}%
                                   {-3.25ex\@plus -1ex \@minus -.2ex}%
                                   {1.5ex \@plus .2ex}%
                                   {\normalfont\normalsize\bfseries}}
\renewcommand\subsubsection{\@startsection{subsubsection}{3}{\z@}%
                                   {-3.25ex\@plus -1ex \@minus -.2ex}%
                                   {1.5ex \@plus .2ex}%
                                   {\normalfont\normalsize\it}}
\renewcommand\paragraph{\@startsection{paragraph}{4}{\z@}%
                                   {-3.25ex\@plus -1ex \@minus -.2ex}%
                                   {1.5ex \@plus .2ex}%
                                   {\normalfont\normalsize\bf}}

\numberwithin{equation}{section}

\def\ie{{\it i.e.}}
\def\eg{{\it e.g.}}

\def\revise#1       {\raisebox{-0em}{\rule{3pt}{1em}}%
                     \marginpar{\raisebox{.5em}{\vrule width3pt\
                     \vrule width0pt height 0pt depth0.5em
                     \hbox to 0cm{\hspace{0cm}{%
                     \parbox[t]{4em}{\raggedright\footnotesize{#1}}}\hss}}}}

\def\cala         {{\cal A}}

\def\cald         {{\cal D}}
\def\cale         {{\cal E}}
\def\calf         {{\cal F}}

\def\call         {{\cal L}}
\def\calm         {{\cal M}}
\def\caln         {{\cal N}}
\def\calo         {{\cal O}}
\def\calp         {{\cal P}}

\def\del          {\partial}

\def\ee           {{\rm e}}

\def\tr           {\mathop{\rm Tr}}

\def\sqr#1#2{{\vcenter{\vbox{\hrule height.#2pt
 \hbox{\vrule width.#2pt height#1pt \kern#1pt
 \vrule width.#2pt}\hrule height.#2pt}}}}

\newcommand{\fft}[2]{{\frac{#1}{#2}}}
\newcommand{\ft}[2]{{\textstyle{\frac{#1}{#2}}}}
\def\jsquare{\mathop{\mathchoice{\sqr{8}{32}}{\sqr{8}{32}}
{\sqr{6.3}{9}}{\sqr{4.5}{9}}}}

\def\a{\alpha}
\def\w{\omega}

\def\a{\alpha}
\def\r{\rho}

\def\rh{\hat{\rho}}
\def\chih{\hat{\chi}}
\def\dd{\delta}

\def\ee{\epsilon}
\def\q{\bar{q}}
\def\<{\langle}
\def\>{\rangle}

\catcode`\@=12

\begin{document}


\title{$\caln=2^*$ hydrodynamics}

\pubnum{%
hep-th/0406200}
\date{June 2004}

\author{
Alex Buchel\\[0.4cm]
\it Perimeter Institute for Theoretical Physics\\
\it Waterloo, Ontario N2J 2W9, Canada\\[0.2cm]
\it Department of Applied Mathematics\\
\it University of Western Ontario\\
\it London, Ontario N6A 5B7, Canada\\
}

\Abstract{
Using gauge theory /string theory correspondence certain universal
aspects of the strongly coupled four dimensional gauge theory
hydrodynamics were established in hep-th/0311175. The analysis were
performed in the framework of ``membrane paradigm'' approach to the
fluctuations on the black brane stretched horizon.  We confirm the
universal result for the shear viscosity to the entropy density ratio
for the strongly coupled $\caln=2^*$ gauge theory from explicit
computation of the finite temperature Minkowski-space correlation
functions in the dual supergravity geometry.
}


\makepapertitle

\body

\version\versionno

\section{Introduction}
Gauge theory/string theory  correspondence \cite{m9711} represents a valuable tool for 
analyzing  dynamics of gauge theories. 
In particular, it provides an  effective description of finite temperature strongly coupled gauge
theories in terms of supergravity black hole backgrounds. Recently, with the formulation of the 
prescription for the computation of Minkowski-space correlation functions 
in gauge/gravity correspondence \cite{ss,hs}, a study of {\it non-equilibrium} processes 
(\eg\ diffusion and sound propagation) in gauge theory plasma became possible \cite{ne1,ne2,ne3,ne4,ne5}.   
There is a  hope  that dual supergravity  analysis of plasma transport coefficients 
will be  of utility in hydrodynamic models used to describe elliptic flows in heavy ion collision experiments 
at RHIC \cite{r1,r2,r3}. 

From the fundamental perspective, an intriguing spin-off of the strongly coupled gauge 
theory hydrodynamics analysis was the formulation by Kovtun, Son and Starinets (KSS) \cite{kss}   
the gauge  theory shear viscosity $\eta$ bound in terms of its entropy density $s$
\begin{equation}
\frac{\eta}{s}\ge \frac{\hbar}{4\pi k_B}\approx 6.08\ \times\ 10^{-13}\ {\rm K\cdot s}\,.
\eqlabel{bound}
\end{equation}  
Specifically, in the framework of ``membrane paradigm'' \cite{p1,p2}
approach to the fluctuations on black brane stretched horizon, KSS established
saturation of the viscosity bound  \eqref{bound} for all maximally supersymmetric gauge theories 
and for $\caln=2^*$ gauge  theory (to leading order in $m/T$) \cite{pw,bl}. In the case of 
maximally supersymmetric gauge theories, ``membrane paradigm'' computations were shown to 
reproduce analysis of the transport coefficients extracted from Minkowski-space correlators.     
Moreover, since the near horizon black brane geometries dual to finite temperature maximally supersymmetric
gauge theories allow for an extension to asymptotically flat space-times, the universality of 
$\eta/s$ can be related  \cite{kss1,bm} to  the universality of low energy absorption cross sections for 
black holes observed in \cite{das}. We emphasize that existence of the asymptotically flat region is 
absolutely crucial to establish the latter connection, as only in that case one can set up a 
graviton scattering ``experiment'' of \cite{das}.

Membrane paradigm framework toward gauge theory hydrodynamics \cite{kss}, though lacking a rigorous 
physical understanding enjoyed within correlation function approach \cite{ne2}, is ultimately 
the most flexible (and the easiest to implement). For one reason, it does not 
rely on the existence of the asymptotically flat region\footnote{Unlike 
maximally supersymmetric examples of the gauge/gravity correspondence, asymptotically flat 
extensions of the supergravity backgrounds dual to non-conformal gauge theories are not known.}.
Second, such computations are sensitive only to the local geometry in the vicinity of the 
horizon, and thus following the  gauge theory/ string theory correspondence 
should describe the {\it infrared} properties of the dual gauge theory. This fits neatly 
with the viewpoint of hydrodynamics as a {\it low-energy} effective description of the 
system close to equilibrium. Also, this makes an observation that supergravity typically 
realize standard four-dimensional non-conformal gauge theory only in the IR 
(as in \cite{ks,mn}) irrelevant. In fact, correlation function approach toward computation 
of the transport coefficients for the gauge theories dual to geometries of \cite{ks,mn}
is rather subtle, as it is linked to (yet unresolved) issues of local regularization 
of such models \cite{ab}. Within membrane paradigm approach, a theorem was proved 
\cite{bl1} that all systems, admitting a supergravity holographic dual 
(with a low-energy gauge theory description) saturate \eqref{bound}
at infinite 't Hooft coupling\footnote{Finite 't Hooft coupling corrections 
(dual to $\alpha'$-corrections on the string theory side of the correspondence) 
to \eqref{bs} are nonvanishing, and will be reported in \cite{bls}.
}:
\begin{equation}
\begin{split}
&\frac{\eta}{s}=f\left(g_{YN}^2N, \frac{\Lambda_i}{T}\right)\ \frac{1}{4\pi}\,, \\
&\lim_{g_{YN}^2N\to\infty} f\left(g_{YN}^2N, \frac{\Lambda_i}{T}\right)=1\,,
\end{split}
\eqlabel{bs}
\end{equation}  
independent of any microscopic scales $\{\Lambda_i\}$ relative to the temperature $T$.
 
Given a somewhat surprising conclusion \eqref{bs}, and a conjectural status of 
the membrane paradigm approach, we feel 
independent verification of \eqref{bs} for non-conformal gauge theories is highly desirable. 
Such a verification is provided here via explicit computation of the Minkowski-space 
correlation functions in the supergravity dual to finite temperature $\caln=2^*$ 
gauge theory. Though the relevant black hole geometry \cite{bl} is not known 
analytically, we will be able to perform analytical analysis of the shear 
mode fluctuations using two complimentary approaches. First, we compute 
shear viscosity using Kubo formula from the  correlation function of 
the stress-energy tensor at zero spatial momentum
\begin{equation}
\eta=\lim_{\w\to 0}\ \frac{1}{2\w}\int dtd\bar{x}\ e^{i\w t}\ \langle
[T_{xy}(x),T_{xy}(0)]\rangle\,.
\eqlabel{kubo}
\end{equation}    
We find that though 
\begin{equation}
\eta=\eta\left(T,\frac{m_b}{T},\frac{m_f}{T}\right)\,,    
\eqlabel{en2}
\end{equation} 
where \cite{bl} $m_b$  ($m_f$) are (generically different) masses of the 
bosonic (fermionic) components of the $\caln=2^*$ hypermultiplet, 
\begin{equation}
\frac{\eta}{s}=\frac{1}{4\pi}\,,
\eqlabel{res1}
\end{equation}
independent of the conformal symmetry breaking scales $m_b,\ m_f$. 
Second, from the stress-tensor correlation functions which have a diffusion pole, 
we directly extract the shear diffusion constant $\cald$
\begin{equation}
\cald=\frac{1}{4\pi T}\,,
\eqlabel{diff}
\end{equation} 
which given identity 
\begin{equation}
\frac{\eta}{s}=\cald T\,,
\eqlabel{id}
\end{equation}
reproduces Kubo formula result \eqref{res1}.

Of cause, prior to computation of the correlation functions, 
the thermodynamics of the supergravity background must be understood. 
While the non-extremal deformation of the PW flow \cite{pw} was 
constructed in \cite{bl}, the complete understanding of its thermodynamics 
was lacking. Using \cite{ab}, we resolve the puzzle of the black hole thermodynamics of 
\cite{bl}. 

\section{$\caln=2^*$ thermodynamics}

\subsection{The geometry}
The supergravity background dual to finite temperature $\caln=2^*$ gauge 
theory was studied in  \cite{bl}. Here we collect the relevant facts about 
the geometry referring for the details to the original analysis.

The effective five-dimensional action is
\begin{equation}
\begin{split}
S=&\,
\int_{\calm_5} d\xi^5 \sqrt{-g}\ \call_5\\
=&\frac{1}{4\pi G_5}\,
\int_{\calm_5} d\xi^5 \sqrt{-g}\left(\ft14 R-3 (\del\a)^2-(\del\chi)^2-
\calp\right)\,,
\end{split}
\eqlabel{action5}
\end{equation}
where the potential $\calp$ is%
\footnote{We set the 5d gauged supergravity coupling to one. This
corresponds to setting the $S^5$ radius $L=2$.}
\begin{equation}
\calp=\frac{1}{16}\left[\frac 13 \left(\frac{\del W}{\del \a}\right)^2+
\left(\frac{\del W}{\del \chi}\right)^2\right]-\frac 13 W^2\,,
\eqlabel{pp}
\end{equation}
with the superpotential
\begin{equation}
W=-\frac{1}{\r^2}-\frac 12 \r^4 \cosh(2\chi)\,.
\eqlabel{supp}
\end{equation}
The five dimensional Newton's constant is 
\begin{equation}
G_5\equiv \frac{G_{10}}{2^5\ {\rm vol}_{S^5}}=\frac{4\pi}{N^2}\,.
\eqlabel{g5}
\end{equation}
The action \eqref{action5} yields the Einstein equation
\begin{equation}
\frac 14 R_{\mu\nu}=3\del_\mu \a \del_\nu\a+\del_\mu\chi\del_\nu\chi
+\frac13 g_{\mu\nu} \calp\,,
\eqlabel{ee}
\end{equation}
and the scalar equations
\begin{equation}
\jsquare\alpha=\fft16\fft{\del\calp}{\del\alpha}\,,\qquad
\jsquare\chi=\fft12\fft{\del\calp}{\del\chi}\,.
\eqlabel{scalar}
\end{equation}
For a finite temperature deformation of the PW \cite{pw} flow metric
we take
\begin{equation}
ds_5^2=e^{2 A} \left(-e^{2 B}\ dt^2 +d\vec{x}\,^2\right)+dr^2\,,\\
\eqlabel{ab}
\end{equation}
where $e^{2B}$ represents a blackening function.  Note that we choose to
retain $g_{rr}=1$ since any non-trivial factor can be absorbed into a
redefinition of $r$.

Substituting this metric ansatz into the equations of motion,
\eqref{ee} and \eqref{scalar}, we find
\begin{equation}
\begin{split}
0&=\a''+\left(4 A' + B'\right)\a' -\frac 16 \frac{\del\calp}{\del\a}\,,\\
0&=\chi''+\left(4 A' + B'\right)\chi' -\frac 12 \frac{\del\calp}{\del\chi}
\,,\\
0&=B''+\left(4 A'+B'\right) B'\,,\\
&\frac 14 A''+\frac 14 B''+
\left(A'\right)^2+\frac 14\left(B'\right)^2+\frac 54 A' B'=-\frac 13 \calp\,,\\
&- A''-\frac 14  B''-\left(A'\right)^2
-\frac 14 \left(B'\right)^2-\frac 12 A' B'
=3\left(\a'\right)^2+\left(\chi'\right)^2 +\frac 13\calp\,.
\end{split}
\eqlabel{aeq}
\end{equation}
Notice that the equation for $B$ in \eqref{aeq} can be integrated once to obtain
\begin{equation}
\ln B' +4 A +B\ =\ {\rm const}\,.
\eqlabel{intB}
\end{equation}
This relation will prove useful below.

Nonsingular in the IR flows of \eqref{aeq} are given by a three parameter family
$\{\alpha,\r_0>0,\chi_0\}$, specifying the near horizon ($r\to0$)
Taylor series expansions
\begin{equation}
\begin{split}
e^A&=e^{\a}\,
\left[1+\left(\sum_{i=1}^{\infty}\ a_i\ r^{2 i}\right)\right]\,,\\
e^B&=\delta\ r\left(1+\sum_{i=1}^{\infty}\ b_i\ r^{2 i} \right)\,,\\
\r&=\r_0+\left(\sum_{i=1}^{\infty}\ \r_i\ r^{2 i}\right)\,,\\
\chi&=\chi_0+\left(\sum_{i=1}^{\infty}\ \chi_i\ r^{2 i}\right)\,.
\end{split}
\eqlabel{ds0asa}
\end{equation}
Here, $\delta=\delta(\r_0,\chi_0)$ should be adjusted so that
$e^B\to 1_-$ as $r\to +\infty$.  The first non-trivial terms in the
series expansions \eqref{ds0asa} are
\begin{equation}
\begin{split}
\dd^{-2}\ a_1&=\ft{1}{12}\ \r_0^{-4}+\ft{1}{6}\ \r_0^2\ \cosh(2\chi_0)-
\ft{1}{48}\ \r_0^8\ \sinh^2(2\chi_0)\,,\\
\dd^{-2}\ b_1&=-\ft{1}{9}\ \r_0^{-4}-\ft{2}{9}\ \r_0^2\ \cosh(2\chi_0)+
\ft{1}{36}\ \r_0^8\ \sinh^2(2\chi_0)\,,\\
\dd^{-2}\ \r_1&=\ft{1}{24}\ \r_0^{-3}-\ft{1}{24}\ \r_0^3\ \cosh(2\chi_0)+
\ft{1}{48}\ \r_0^9\ \sinh^2(2\chi_0)\,,\\
\dd^{-2}\ \chi_1&=-\ft{1}{8}\ \r_0^{2} \sinh(2\chi_0)
    +\ft{1}{64}\ \r_0^8\ \sinh(4\chi_0)\,.
\end{split}
\eqlabel{1dsa}
\end{equation}
Three integration constants
$\{\a,\chi_0,\rho_0\}$  are related to temperature and masses of the
$\caln=2$ hypermultiplet components. The most
general solution of \eqref{aeq} in the UV ($\chi\to 0_+$) has altogether
five parameters, $\{\xi,\rh_{10},\rh_{11},\chih_0,\chih_{10}\}$.
Three of them are related to the temperature and the masses,
while the other two are uniquely determined from the requirement of having
a regular horizon, \eqref{1dsa}.  In any case, we have a three parameter
BH solution%
\footnote{The integration constant $\beta$ can be absorbed at the expense
of shifting the position of the horizon in the radial coordinate $r$,
or alternatively by rescaling $x$.}
\begin{equation}
\begin{split}
B\sim\,&-\beta\,x^4\bigl[1+\ft89 x^2 \chih_0^2+x^4\bigl(
\ft{5}{16}\rh_{11}^2-\ft12\rh_{11}\rh_{10}+\ft{1}{18}
\chih_0^4+2\rh_{10}^2+\chih_{0}^2\chih_{10}\\
&\qquad+\ln x \left(-\ft12 \rh_{11}^2+\ft43\chih_0^4+4\rh_{11}\rh_{10}
\right)+2\rh_{11}^2 \ln^2 x\bigr)\bigr]\,,\\
\end{split}
\eqlabel{ex1}
\end{equation}
\begin{equation}
\begin{split}
\chi\sim\,&\chih_0\,x\,\bigl[
1+x^2 \left(\chih_{10}+\ft43 \chih_0^2\ \ln x \right)
+x^4\bigl(\ft{31}{8} \rh_{11}^2-\ft{13}{2}\rh_{11}\rh_{10}
-\ft{56}{45}\chih_0^4-\ft32 \chih_0^2\rh_{11}+2\chih_0^2\rh_{10}\\
&\qquad+5 \rh_{10}^2+2\chih_0^2\chih_{10}
+\ln x\left(-\ft{13}{2}\rh_{11}^2+10\rh_{11}\rh_{10}+\ft83
\chih_0^4+2\chih_0^2\rh_{11}\right)+5\rh_{11}^2\ln^2 x\bigr)\bigr]\,,\\
\end{split}
\eqlabel{ex2}
\end{equation}
\begin{equation}
\begin{split}
\r\sim\,&1+x^2\left(\rh_{10}+\rh_{11}\ln x\right)
+x^4\bigl(-2 \rh_{11}\rh_{10}+\ft32 \rh_{11}^2+\ft32\rh_{10}^2
+\ft{10}{3}\chih_0^2\rh_{10}-\ft{8}{3}\chih_0^2\rh_{11}+\ft13\chih_0^4\\
&\qquad+\ln x \left(3\rh_{11}\rh_{10}+\ft{10}{3}
\chih_0^2\rh_{11}-2\rh_{11}^2\right)+\ft32 \rh_{11}^2\ln^2 x\bigr)\,,\\
\end{split}
\eqlabel{ex3}
\end{equation}
\begin{equation}
\begin{split}
A\sim\,&\xi-\ln x -\ft 13 \chih_{0}^2 x^2+x^4\bigl(\ft14\beta
+\ft19 \chih_0^4-\ft12\chih_0^2\chih_{10}-\ft{1}{8}\rh_{11}^2-\rh_{10}^2\\
&\qquad-\ln x \left(\ft23 \chih_0^4+2\rh_{11}\rh_{10}\right)
-\rh_{11}^2\ln^2 x\bigr)\,.
\end{split}
\eqlabel{ex5}
\end{equation}
Also, we find
\begin{equation}
\begin{split}
\frac{dA}{dr}\sim\ & \ft12 +\ft13 \chih_0^2 x^2+x^4
\bigl(-\ft12\beta +2 \rh_{10}^2+\rh_{11}\rh_{10}
+\ft14 \rh_{11}^2+\chih_0^2\chih_{10}+\ft{1}{9}\chih_{0}^4\\
&+\ln x \left(\rh_{11}^2 +\ft43 \chih_{0}^4+4\rh_{11}\rh_{10}
\right)+2\rh_{11}^2\ln^2 x\bigr)\,.
\end{split}
\eqlabel{ex6}
\end{equation}
In \eqref{ex1}-\eqref{ex6}, $x=x_0 e^{-r/2}$ with $x_0$ an
arbitrary constant.

It is straightforward to compute the Bekenstein-Hawking entropy density 
\begin{equation}
s=\frac {\cala_{horizon}}{4 G_N}=
\frac 12\,\pi^2 N^2\,\left(
\frac{1}{2\pi}\ e^{\a}\right)^3\,,
\eqlabel{entropy}
\end{equation}
and the black hole temperature
\begin{equation}
T=\frac{1}{2\pi} e^{A} \left(\frac{\del e^B}{\del r}\right)\bigg|_{r\to 0_+}
\!\!=\frac{1}{2\pi} e^{\a}\ \dd\,.
\eqlabel{Tdef}
\end{equation}

\subsection{Boundary renormalization}
Understanding the black hole thermodynamics mandates understanding the 
corresponding geometry energy (ADM mass) and  the free energy. 
Following ideas of \cite{re1,re2,re3}, the ADM mass
can be computed as a one-point correlation function of the 
boundary stress tensor, while  the free energy  can be extracted from the 
expectation value of the Euclidean gravitation  action. 
Both quantities are infinite and must be properly regularized and 
renormalized\footnote{Understanding the renormalization of the gravitation action is 
also vital for the computation of the correlation functions in the framework of gauge theory 
/ string theory correspondence.}.
Below, we present the necessary results while referring for details to \cite{ab}. 

Let $r$ be the position of the boundary, and $S_E^r$ be the Euclidean gravitational action 
 on the cut-off space 
\begin{equation}
\lim_{r\to \infty} S_E^r= S_E\,,
\eqlabel{cutac}
\end{equation}
where  $S_E$ is the Euclidean version of \eqref{action5}.
Besides the standard Gibbons-Hawking term 
\begin{equation}
S_{GH} = -\frac{1}{8\pi G_5} \int_{\del\calm_5} d^4x \sqrt{h_E}
\nabla_{\mu} n^{\mu} = -\frac{1}{8\pi G_5} \bigg[\left(e^{4A+3 B}\right)'
\bigg]\bigg|^r\int_{\del\calm_5}d^4 x\,,
\eqlabel{gh}
\end{equation}
we supplement the combined 
regularized action $\left(S_E^r+S_{GH}\right)$ by the appropriate boundary 
counterterms which are needed to get a finite action. These boundary 
counterterms must be constructed from the local
metric and $\{\a=\ln\r,\chi\}$ scalar invariants on the boundary $\del\calm_5$,
except for the terms associated with the conformal anomaly  which
include an explicit dependence on the position of the boundary,
\begin{equation}
\begin{split}
S^{counter}&=\frac{1}{4\pi G_5}\int_{\del\calm_5}d^4x \sqrt{h_E}\biggl(
\a_1+\a_2\ R_4
+\a_3\ \a+\a_4\ \chi^2\cr &+\a_5\ \a^2+\a_6\ \a\chi^2+\a_7\ R_4\ \a+
\a_8\ \frac{\a^2}{\ln \ee}+ \ln \ee (\a_9 R_4\ \chi^2 + \a_{10} \chi^4)
+\a_{11}\chi^4\biggr).
\end{split}
\eqlabel{scount1}
\end{equation}
Here $R_4\equiv R_4(h_E)$ is the  Ricci scalar constructed from $h_{\mu\nu}$
and the
$\a_i$ are constant coefficients of the counterterms which are determined
by the requirement of having a finite action (the coefficients $\a_8,
\a_9, \a_{10}$ which are associated with the conformal anomaly have been
computed before  but we leave them arbitrary for now).
The quartic term in $\chi$ is actually finite and is 
introduced to preserve supersymmetry.
The conformal anomaly terms depend on the position of the boundary; we
choose to parameterize this position by the physical quantity \cite{ab}
\begin{equation}
\ee\equiv \sqrt{-g_{tt}}\,.
\eqlabel{eedef}
\end{equation} 
 The counterterm parameters  $\a_i$ are fixed in such a way 
that the {\it renormalized} Euclidean action $I_E$ is finite 
\begin{equation}
I_E\equiv \lim_{r\to \infty}\ \biggl(
S_E^r+S_{GH}+S^{counter}\biggr)\,,\qquad |I_E|<\infty\,.
\eqlabel{IEdef1}
\end{equation}
We find \cite{ab} 
\begin{equation}
\begin{split}
\a_1&=\frac 34\,,\qquad \a_2=\frac{1}{4}\,,\qquad \a_3=0\,,\qquad 
\a_4=\frac 12\,,\qquad
\a_5=3\,,\qquad \a_6=0\,,\\
\a_7&=0\,,\qquad \a_8=-\frac 32\,,
\qquad \a_9=- \frac{1}{3}\,,\qquad \a_{10}=-\frac 23\,,\qquad  \a_{11}=\frac 16\,.
\end{split}
\eqlabel{ai1}
\end{equation}

The quasilocal stress tensor $T_{\mu\nu}$ for our background
is obtained from the variation of the full action 
\begin{equation}
S_{tot}=S_E^r+S_{GH}+S^{counter}\,,
\eqlabel{total1}
\end{equation}
with respect to the boundary metric $\delta h_{\mu\nu}$
\begin{equation}
T^{\mu\nu}=\frac{2}{\sqrt{-h}}\ \frac{\delta S_{tot}}{\delta h_{\mu\nu}}\,.
\eqlabel{qlst1}
\end{equation}  
Explicit computation yields 
\begin{equation}
\begin{split}
T^{\mu\nu}=&\frac{1}{8\pi G_5}\biggl[
-\Theta^{\mu\nu}+\Theta h^{\mu\nu}
\\
&-2\left\{\a_1
+\a_3\ \a+\a_4\ \chi^2+\a_5\ \a^2+\a_6\ \a\chi^2
+\a_8\ \frac{\a^2}{\ln \ee}+\a_{10}\ \ln\ee\ \chi^4+\a_{11}\ \chi^4
\right\} h^{\mu\nu}\\
&+4\left\{
\a_2+\a_7\ \a+\a_9\ \ln \ee\ \chi^2\right\}
 \left(R_4^{\mu\nu}-\ft 12 R_4 h^{\mu\nu}\right)\biggr]\,,
\end{split}
\eqlabel{tfin}
\end{equation}
where 
\begin{equation}
\Theta^{\mu\nu}=\ft 12 \left(\nabla^\mu n^\nu+
\nabla^\nu n^\mu\right),\qquad \Theta=\tr \Theta^{\mu\nu}\,.
\eqlabel{thdef}
\end{equation}

\subsection{The thermodynamics}
The thermodynamics of the PW flow \cite{pw} has been studied 
before in \cite{bl}. In \cite{bl} computation of the 
free energy and the energy (ADM mass) has been done using the 
background subtraction prescription of \cite{hh}, with 
a reference background being the supersymmetric PW flow. 
Though one can regularize in this way the free energy $F$,
the energy $E$, prove the relation
\begin{equation}
F=E-T S\,, 
\eqlabel{t1}
\end{equation}
where $S$, $T$ are the entropy and the Hawking temperature of the 
nonextremal PW deformation, one finds 
\begin{equation}
T\ dS\ne dE\,.
\eqlabel{t2}
\end{equation}
The disagreement with the first law of thermodynamics either points to 
incorrect subtraction prescription, or to a more exotic explanation, 
like a chemical potential \cite{bl}. That subtraction 
prescription \cite{hh} does not always work is well known. In fact, 
in some cases one simply can not find an appropriate reference geometry \cite{bp}.
What is surprising with non-extremal deformation of the PW flow, is the fact  that a well-motivated 
reference background leads to a problematic thermodynamics. 
An alternative prescription of regularizing the free energy and the ADM mass
of a gravity background dual to a gauge theory (in a sense of \cite{m9711})
has been formulated  in \cite{re1}, and applied to an example where 
approach of \cite{hh} can not work, in \cite{bp}. 
With  understanding of  the renormalization of the gravity dual to $\caln=2^*$
gauge theory in previous section, we can study its thermodynamics following 
\cite{re1,j1,bp}.  In this section we show that with the counterterm structure \eqref{scount1},
\eqref{ai1} 
the first law  of thermodynamics for the nonextremal deformation of the PW solution \cite{bl} holds.
We follow mostly \cite{bl}. Note the difference in normalization of a 5d scalar $\a$ here and in \cite{bl},
$\a_{BL}$
\begin{equation}
\a_{BL}\equiv \sqrt{3}\a,\qquad \r_{BL}\equiv \r\,.
\eqlabel{normal}
\end{equation}

As in \cite{bl} we identify the regularized Euclidean action with $F/T$
\begin{equation}
F\equiv T I_E\,,
\eqlabel{t3}
\end{equation} 
and the energy (ADM mass) $E$ with 
\begin{equation}
E=\int_{\Sigma}d^3\xi\ \sqrt{\sigma} N_{\Sigma} \cale\,,
\eqlabel{massdef}
\end{equation}
where\footnote{We take the volume 
of $R^3$ to be $V_3$.} $\Sigma\equiv R^3$ is a spacelike hypersurface in $\del\calm_5$
with a timelike unit normal $u^\mu$,
$N_{\Sigma}$ is the norm of the timelike Killing vector 
in \eqref{ab}, ${\sigma}$ is the determinant of the induced metric 
on $\Sigma$, and $\cale$ is the proper energy density 
\begin{equation}
\cale=u^\mu u^\nu T_{\mu\nu}\,.
\eqlabel{epdef}
\end{equation}
The quasilocal stress tensor $T_{\mu\nu}$ is computed following \eqref{qlst1}.
Combining eqs.~(6.18),(6.22) of \cite{bl} with the  relevant terms of 
\eqref{scount1} we find the equivalent of \eqref{IEdef1}
\begin{equation}
\begin{split}
I_E=&\frac{1}{8\pi  G_5}\frac{V_3}{T}\biggl(-e^{3A}\left(e^{A+B}\right)'\bigg|_{horizon}
+\lim_{r\to\infty}\biggl[-3e^{4A+B}A'+2e^{4A+B}\biggl\{\a_1+\a_3\ \a\\
&+\a_4\ \chi^2
+\a_5\ \a^2+\a_6\ \a\chi^2+\a_8\ \frac{\a^2}{\ln\ee}
+\ln\ee\ \a_{10}\ \chi^4+\a_{11}\chi^4
\biggr\}\biggr]\biggr)\\
\equiv& \frac{1}{8\pi  G_5}\frac{V_3}{T}\biggl(-e^{3A}\left(e^{A+B}\right)'\bigg|_{horizon}
+\triangle_{BH}\biggl)\\
=&-S+\frac{V_3\triangle_{BH}}{8\pi G_5 T}\,,
\end{split}
\eqlabel{iregb}
\end{equation}
where $S$ is the entropy of the black hole horizon, 
and we defined $\triangle_{BH}$ as above.
Additionally, as in \eqref{eedef} we identify
\begin{equation}
\ee\equiv \sqrt{-g_{tt}}=e^{A+B}\,.
\end{equation} 
Also 
\begin{equation}
\begin{split}
T_{tt}=&\frac{h_{tt}}{8\pi G_5}\biggl[
-3A'+2\biggl\{\a_1+\a_3\ \a
+\a_4\ \chi^2
+\a_5\ \a^2+\a_6\ \a\chi^2+\a_8\ \frac{\a^2}{\ln\ee}\\
&+\ln\ee\ \a_{10}\ \chi^4+\a_{11}\chi^4
\biggl\}\biggr]\,.
\end{split}
\eqlabel{admr}
\end{equation}
From \eqref{massdef}, \eqref{admr} we have 
\begin{equation}
\begin{split}
E=&\frac{1}{8\pi G_5} V_3\ \lim_{r\to\infty}\biggl[-3e^{4A+B}A'+2e^{4A+B}\biggl\{\a_1+\a_3\ \a\\
&+\a_4\ \chi^2
+\a_5\ \a^2+\a_6\ \a\chi^2+\a_8\ \frac{\a^2}{\ln\ee}
+\ln\ee\ \a_{10}\ \chi^4+\a_{11}\chi^4
\biggr\}\biggr]\\
=&\frac{V_3 \triangle_{BH}}{8\pi G_5}\,.
\end{split}
\eqlabel{efin}
\end{equation}
Notice that from \eqref{t3}, \eqref{iregb}, \eqref{efin} we trivially recover 
\eqref{t1}.
The asymptotic solution for $\{A,B,\r,\chi\}$ is given in \eqref{ex1}-\eqref{ex6},
using which we find that $\triangle_{BH}$ is indeed finite with 
\begin{equation}
\begin{split}
\triangle_{BH}=&-\frac{1}{12} e^{4\xi}\left(-18\beta+9\hat\r_{11}^2+12\hat\chi_0^2\hat\chi_{10}
-36\hat\r_{11}\hat\r_{10}+16\hat\chi_0^4\xi-36\hat\r_{11}^2\xi\right)\,,
\end{split}
\eqlabel{trfin}
\end{equation} 
where we used \eqref{ai1}.
As in \cite{bl}, we can not study analytically the thermodynamics apart from the 
high temperature regime, $\ft mT\ll 1$, where $m$ is the mass of the $\caln=2$ 
hypermultiplet of the dual gauge theory. Using the leading order  analytical 
solution of Sec.~5 of \cite{bl}, specifically eqs.~(6.40)-(6.45), we find
 \begin{equation}
\begin{split}
S&=\frac 12 \pi^2 N^2 T^3   \left(1-
\frac{\Gamma(3/4)^4}{\pi^4}\,\frac{m^2}{T^2}\right)\,,\\
E&=\frac 38 \pi^2 N^2 T^4 \left(1-\frac 23
\frac{\Gamma(3/4)^4}{\pi^4}\,\frac{m^2}{T^2}\right)\,,\\
F&=-\frac 18 \pi^2 N^2 T^4  \left(1- 2
\frac{\Gamma(3/4)^4}{\pi^4}\,\frac{m^2}{T^2}\right)\,.
\end{split}
\eqlabel{thermofin}
\end{equation}
with 
\begin{equation}
T\ dS= dE\,.
\eqlabel{t4}
\end{equation}

\section{The viscosity bound}
In this section we closely follow \cite{ne2}.
To compute two point correlation functions of the  boundary stress-energy tensor
we consider small  fluctuations of the near-extremal PW flow metric 
\begin{equation}
g_{\mu\nu}=g_{\mu\nu}^{(0)}+h_{\mu\nu}\,,
\end{equation}   
where $g_{\mu\nu}^{(0)}$ is the background metric \eqref{ab}. As in \cite{ne2},
we assume that metric perturbations take the form
\begin{equation}
h_{\mu\nu}=e^{-i\w t+i q z} H_{\mu\nu}(r)\,. 
\eqlabel{pmet}
\end{equation} 
Residual $O(2)$ symmetry of rotations in $xy$ plane 
guarantees that (at a linearized level) field equations for 
$\{h_{xy}\}$ and $\{h_{tx},h_{xz}\}$ (as well as the other metric components) 
decouple. We use $h_{xy}$ perturbations for the computation of the $T_{xy}T_{xy}$
correlation functions to be used in Kubo formula \eqref{kubo}; the (coupled)
fluctuations   $\{h_{tx},h_{xz}\}$ will produce correlation functions with a 
diffusion pole.

\subsection{Shear viscosity via Kubo relation}
Eq.~\eqref{kubo} can be written in the form 
 \begin{equation}
\eta = \lim_{\omega \rightarrow 0} {\frac {1}{2 \omega i}} \Biggl[ 
G_{xy,xy}^A (\omega,0) - G_{xy,xy}^R (\omega,0)\Biggr]\,,
\eqlabel{RA}
\end{equation}
where the retarded Green's function is defined as
\begin{equation}
  G_{\mu\nu,\lambda\rho}^R (\omega, \q)
  = -i\!\int\!d^4x\,e^{-iq\cdot x}\,
  \theta(t) \< [T_{\mu\nu}(x),\, T_{\lambda\rho}(0)] \>\,,
\eqlabel{retarded}
\end{equation}
and $G^A(\omega,\q) = (G^R (\omega, \q))^*$.

Following \cite{ss,hs}, retarded correlation function $G_{xy,xy}^R(\w,\q)$ can be extracted 
from the (quadratic) boundary effective action $S_{boundary}$ for the metric fluctuations 
$h_{xy}^b$
\begin{equation}
h_{xy}^b(k)\equiv h_{xy}^b(\w,\q)=\int \frac{d^4k}{(2\pi)^4} e^{-i\w t+i q\cdot x}\
h_{xy}(t,\bar{x},r),\qquad {\rm as}\ \ r\to \infty\,,
\eqlabel{fh}
\end{equation}
given by
\begin{equation}
S_{boundary}[h_{xy}^b]=\int \frac{d^4k}{(2\pi)^4}\ h_{xy}^{b}(-k)\ \calf(k,r)\ h_{xy}^{b}(k)
\bigg|_{r\to 0}^{r\to \infty}\,, 
\eqlabel{sssb}
\end{equation}
as 
\begin{equation}
G^R_{xy,xy}(\w,\q)=\lim_{r\to\infty}\ 8\ \calf(k,r)\,. 
\eqlabel{Gr}
\end{equation}
The boundary metric functional is defined as
\begin{equation}
S_{boundary}[h_{xy}^b]=\lim_{r\to\infty}\biggl(
 S^r[h_{xy}]+S_{GH}[h_{xy}]+S^{counter}[h_{xy}]\biggr)\,,
\eqlabel{bounfu}
\end{equation}
where $S^r$ is the bulk Minkowski-space cut-off action \eqref{ab}, evaluated on-shell 
for the bulk metric fluctuations $h_{xy}(t,\bar{x},r)$ subject to the following boundary conditions:
\begin{equation}
\begin{split}
&(a):\ \lim_{r\to\infty} h_{xy}(t,\bar{x},r)=h_{xy}^b(t,\bar{x})\,,\\
&(b):\ h_{xy}(t,\bar{x},r)\ {\rm  is\ an\ incoming\ wave\ at\ the\ horizon}\ (r\to\infty)\,.  
\end{split}
\eqlabel{bc}
\end{equation}  
The purpose of the boundary counterterm $S^{counter}$ \eqref{scount1} is to remove 
divergent (as $r\to\infty$) and $\{\w,\q\}$-independent contributions from the
kernel $\calf$ of \eqref{sssb}. 
Despite the absence of the analytical solution to \eqref{aeq} for generic bosonic and fermionic 
masses\footnote{Leading in $\{m_b/T,m_f/T\}$ solution was given in \cite{bl}.} 
($m_b$ and $m_f$), we will be able to implement all steps outlined, necessary to 
extract $\eta$ analytically. 

Introducing\footnote{We dropped the spacial dependence in $h_{xy}(t,\bar{x},r)\equiv h_{xy}(t,r)$
as we will need $G^R$ ($G^A$) at zero spacial momentum \eqref{RA}.} 
\begin{equation}
h_{xy}(t,r)\equiv 2\ e^{2A}\ \phi(t,r)\,,
\eqlabel{mcs}
\end{equation}
the effective bulk action for $\phi(t,r)$ becomes that of a minimally 
coupled scalar in the background \eqref{ab}. 
Further introducing a new radial coordinate $y$:
\begin{equation}
y\equiv e^{B(r)}\,,
\eqlabel{ydef}
\end{equation}
so that $y\to 0_+$ corresponds to the horizon, and $y\to 1_-$ corresponds to the boundary, and 
\begin{equation}
\phi(t,r)=e^{-i\w t}\ \phi_k(y)\,,
\end{equation}
we find 
\begin{equation}
\begin{split}
0=&y^2\left(1+4\r^6\cosh^2\chi-2\r^6-\r^{12}\cosh^4\chi+\r^{12}\cosh^2\chi\right)\left(y\ \phi''_k
+\phi'_k\right)\\
&+2e^{-2A}\w^2\r^2\left(3\ \r^2\ A'+6\ y\ \r^2\ \left(A'\right)^2
-6\ y\ \left(\r'\right)^2-2\ y\ \r^2\ \left(\chi'\right)^2
\right) \phi_k\,,
\end{split}
\eqlabel{eomfk}
\end{equation}
where $\{A,\r,\chi\}$ are functions of $y$ satisfying equations equivalent to \eqref{aeq},
also all derivatives are with respect to $y$. 
A low-frequency solution of \eqref{eomfk} which is an incoming wave at the 
horizon, and which near the boundary satisfies 
\begin{equation}
\lim_{y\to 1_- }\phi_k(y)= 1\,,
\eqlabel{bbb}
\end{equation}
can be written as 
\begin{equation}
\phi_k(y)=y^{-i\w Q}\ \biggl(F_0(y)+i\w Q\ F_{\w}(y)+\calo(\w^2)\biggr)\,,
\eqlabel{taylorsol}
\end{equation} 
where, given the Taylor series expansion near the horizon \eqref{ds0asa}, 
\begin{equation}
Q=\frac {1}{\dd}\ e^{-\a}\,,
\eqlabel{defq}
\end{equation}
and   $\{F_0,F_{\w}\}$ satisfy very simple ODE's:
\begin{equation}
\begin{split}
0=&F_0'+y\ F_0''\,,\\
0=&y\ F_\w''+F_\w'-2\ F_0'\,.
\end{split}
\eqlabel{difff}
\end{equation}
The only nonsingular solution to \eqref{difff} which also  
 satisfies \eqref{bbb} is  
\begin{equation}
F_0(y)= 1\,,\qquad F_{\w}(y)= 0\,.
\eqlabel{solf}
\end{equation}
Thus, 
\begin{equation}
\phi(t,r)=e^{-i\w t}\ e^{-i\w Q B(r)}\ \left(1+\calo(\w^2)\right)\,.
\eqlabel{solff}
\end{equation}

Once the bulk fluctuations are on-shell (\ie,\ satisfy  equations of motion)
the bulk gravitational Lagrangian becomes a total derivative. Indeed, 
we find (without dropping any terms)
\begin{equation}
16 \pi G_5\ \call_5=\del_t J^t+\del_rJ^r\,,
\eqlabel{tder}
\end{equation}
where 
\begin{equation}
\begin{split}
J^r=&\frac 32 e^{4A(r)+B(r)}\ 
\phi(t,r)\frac{\del\phi(t,r)}{\del r}+e^{4 A(r)+B(r)}\frac{\del A(r)}{\del r}\ \phi(t,r)^2\,,\\
J^t=&-\frac 32 e^{2 A(r)-B(r)}\ \phi(t,r)\frac{\del\phi(t,r)}{\del t}\,.
\end{split}
\eqlabel{jr}
\end{equation}
Additionally, the Gibbons-Hawking term provides an extra contribution so that 
\begin{equation}
J^r\to J^r-2 e^{4A(r)+B(r)}\ 
\phi(t,r)\frac{\del\phi(t,r)}{\del r}\,.
\eqlabel{ghshift}
\end{equation}
Given the general asymptotic solution \eqref{ex1}-\eqref{ex6}, we find 
\begin{equation}
G^R_{xy,xy}(\w,0)=-\frac{i\w Q e^{4\xi}\beta}{8\pi G_5}+\calo(\w^2)\,.
\eqlabel{grfin}
\end{equation}
Eq.~\eqref{intB} implies (see also eq.(6.33) of \cite{bl})
\begin{equation}
\ln\beta+\ln 2+4\xi=4\a+\ln\dd\equiv \kappa\,.
\eqlabel{cccc}
\end{equation}
Combining \eqref{g5},\eqref{defq} and \eqref{cccc} we rewrite \eqref{grfin} as 
\begin{equation}
G^R_{xy,xy}(\w,0)=-\frac{i\w N^2 e^{3\a}}{64\pi^2}+\calo(\w^2)\,,
\eqlabel{grfin1}
\end{equation}
which leads to \eqref{RA}
\begin{equation}
\eta=\frac{  N^2 e^{3\a}}{64\pi^2}=\frac{1}{4\pi}\ s\,,
\eqlabel{ffinal}
\end{equation}
where we recalled the expression for the entropy density \eqref{entropy}.
As claimed, \eqref{ffinal} reproduces the saturation \eqref{bs} for arbitrary $\{m_b/T,m_f/T\}$.

\subsection{The shear diffusion pole}
Introducing 
\begin{equation}
\begin{split}
h_{tx}(t,z,y)=&e^{-i\w t+i q z}\ e^{2A(y)}\ H_t(y)\,,\\
h_{xz}(t,z,y)=&e^{-i\w t+i q z}\ e^{2A(y)}\ H_z(y)\,,
\end{split}
\eqlabel{po1}
\end{equation}
the coupled system of equations of motion for the diffusive pole channel metric 
fluctuations becomes
\begin{equation}
\begin{split}
0=&H_t'+\frac q\w\ y^2\ H_z'\,,\\
0=&e^{-6 A+2\kappa} \left(y\ H_t''-H_t'\right)-y\left(\w q\ H_z+q^2\ H_t\right)\,,\\
0=&e^{-6 A+2\kappa}\ y\ \left(y\ H_z''+H_z'\right)+\w q\ H_t+\w^2\ H_z\,,  
\end{split}
\eqlabel{eom2}
\end{equation}
where all derivatives are with respect to $y$ defined in \eqref{ydef}, and $\kappa$
is an integration constant \eqref{cccc}.
Solving for $H_z$ from the second equation in \eqref{eom2} and substituting the result 
into the first equation of \eqref{eom2} we find 
\begin{equation}
\begin{split}
0=&e^{-6A+2\kappa} y^2\ G''-e^{-6A+2\kappa} y \left(6y\ A'+2i \w Q-1\right)\ G'\\
&+\left(\w^2+6i\w Q\ y\ e^{-6A+2\kappa}\ A'-\w^2Q^2e^{-6A+2\kappa}-y^2 q^2\right)\ G\,,
\end{split}
\eqlabel{eqg}
\end{equation}
where we further extracted the singular part of $H_t'$ which corresponds 
to the incoming wave boundary condition at the horizon
\begin{equation}
H_t'=y^{-i\w Q+1}\ G(y)\,.
\eqlabel{singlpart}
\end{equation}
Quite amazingly, analytical (regular as $y\to 0_+$) solution 
to \eqref{eqg} to leading order in $\{\w,q^2\}$,
can be found. Specifically, 
\begin{equation}
G(y)\equiv G_0(y)+i\w Q\ G_\w(y)+q^2\ G_q(y)+\calo(\w^2,\w q^2,q^4)\,,
\eqlabel{dddgg}
\end{equation}
with
\begin{equation}
\begin{split}
G_0(y)=&C\,,\\
G_\w(y)=&C\ \int_0^y\ dx\ \frac{1-e^{6A(x)-6\a}}{x}\,,\\
G_q(y)=&\frac C2 \int_0^y\ dx\ x\ e^{6A(x)-2\kappa}\,,
\end{split}
\eqlabel{solsss}
\end{equation}
where we used \eqref{defq} and an explicit expression for $\kappa$ \eqref{cccc}.
Up to an integration constant $C$, solution \eqref{solsss} is unique.
Taking the limit $y\to 1_-$ of the second equation in \eqref{eom2}, and using the 
general asymptotic \eqref{ex5} we find $C$ in terms of the boundary values $H_t^0$ 
and $H_z^0$:
\begin{equation}
C=-\frac{q^2 H_t^0+\w q H_z^0}{i\w Q e^{2\kappa-6\a}-\frac 12 q^2}\,.
\eqlabel{solc}
\end{equation}    
Notice that there is a pole in \eqref{solc} at 
\begin{equation}
i\w=\cald q^2\,,
\eqlabel{polfi}
\end{equation}
with 
\begin{equation}
\cald=\frac {1}{2\dd} e^{-\a}=\frac{1}{4\pi T}\,,
\eqlabel{fincald}
\end{equation}
where we used \eqref{Tdef}. 
As in \cite{ne2}, it is easy to show that the pole in $C$ would result 
in the identical  pole in the retarded correlation functions $G^R_{tx,tx}$,
$G^R_{tx,xz}$, $G^R_{xz,xz}$.

\section{Conclusion}
We performed explicit computations of the Minkowski-space stress-tensor correlation 
functions for $\caln=2^*$ gauge theories at infinite 't Hooft coupling and confirmed 
the universality result \cite{bl}, proved withing membrane paradigm approach.
These computations provide additional support that the KSS bound is saturated 
at infinite 't Hooft coupling irrespectively of the microscopic scales of the system. 
It is an interesting  question as to the modification (if any) of \eqref{bs}
once nonvanishing thermodynamic parameters (\eg,\ chemical potentials) are introduced.
A large class of  supergravity backgrounds relevant for such a study is known.

\section*{Acknowledgments}
I am grateful to  Andrei Starinets for interesting me in the subject, and numerous 
discussions and explanations without which this study would not be possible. 
I would also like to thank Ofer Aharony and Jim Liu for collaborations on 
relevant projects. Research at  Perimeter Institute is supported in part 
by funds from NSERC of Canada. Further support by an 
NSERC Discovery grant is gratefully acknowledged.

\end{document}